\documentclass[twocolumn,secnumarabic,superscriptaddress,amssymb,nobibnotes,aps,prc]{revtex4}
\usepackage{graphicx}
\usepackage{dcolumn}
        
\begin{document} 
\def\Journal#1#2#3#4{{#1} {\bf #2}, #3 (#4)} 
\def\ARNPS{\rm Ann. Rev. Nucl. Part. Sci.}
\def\PPNP{\rm Prog. Part. Nucl. Phys.}
\def\IJMP{\rm. Int. J. Mod. Phys.}
\def\PR{\rm Phys. Rept.}
\def\NCA{\rm Nuovo Cimento}
\def\NCL{\rm Lett. Nuovo Cimento}
\def\NIM{\rm Nucl. Instrum. Methods}
\def\NIMA{{\rm Nucl. Instrum. Methods} A}
\def\NPA{{\rm Nucl. Phys.} A}
\def\NPB{{\rm Nucl. Phys.} B}
\def\PLB{{\rm Phys. Lett.}  B}
\def\PL{{\rm Phys. Lett.}  }
\def\PRL{\rm Phys. Rev. Lett.}
\def\PRD{{\rm Phys. Rev.} D}
\def\PRC{{\rm Phys. Rev.} C}
\def\EPA{{\rm Eur. Phys. J.} A}
\def\EPC{{\rm Eur. Phys. J.} C}
\def\ZPC{{\rm Z. Phys.} C}
\def\ZPA{{\rm Z. Phys.} A}
\def\JPG{{\rm J. Phys.} G}

\title{Impact of nuclear dependence of $R=\sigma_L/\sigma_T$ on 
antishadowing in nuclear structure functions} 
\author{Vadim Guzey}
\affiliation{Hampton University, Hampton, VA 23668, USA}
\author{Lingyan Zhu}
\affiliation{Hampton University, Hampton, VA 23668, USA}
\author{Cynthia E. Keppel}
\affiliation{Hampton University, Hampton, VA 23668, USA}
\author{M. Eric Christy} 
\affiliation{Hampton University, Hampton, VA 23668, USA}
\author{Dave Gaskell}
\affiliation{Thomas Jefferson National Accelerator Facility, Newport News, 
VA 23606, USA}
\author{Patricia Solvignon}
\affiliation{Thomas Jefferson National Accelerator Facility, Newport News, VA 23606, USA}
\author{Alberto Accardi}
\affiliation{Hampton University, Hampton, VA 23668, USA}
\affiliation{Thomas Jefferson National Accelerator Facility, Newport News, 
VA 23606, USA}
\date{\today} 

\begin{abstract}

We study the impact of the nuclear dependence of $R=\sigma_L/\sigma_T$ 
on the extraction of the $F_2^A/F_2^D$ and $F_1^A/F_1^D$ structure function ratios 
from the data on the $\sigma^A/\sigma^D$ cross section ratios. 
Guided by indications of  the nuclear dependence of $R$ from the world data, 
we examine selected sets of EMC, BCDMS, NMC and SLAC data and 
find that $F_1^A/F_1^D < \sigma^A/\sigma^D \leq F_2^A/F_2^D$.
In particular, we observe that the nuclear enhancement (antishadowing) for $F_1^A/F_1^D$ 
in the interval $0.1 < x < 0.3$ 
becomes significantly reduced or even disappears, which  
indicates that antishadowing is dominated by the longitudinal structure function $F_L$.
We also argue that 
precise
measurements of nuclear modifications of $R$ and $F_L^A$ 
have the potential to constrain the poorly known gluon distribution in nuclei 
over
a wide range 
of $x$.

\end{abstract}

\maketitle 

\pagebreak

\section{Introduction}

Since the early lepton scattering experiments  that 
discovered the substructure of the nucleon and eventually led to the development 
of Quantum Chromodynamics (QCD) as the theory of the strong interaction,
deep inelastic scattering (DIS) has been a critical tool in the investigation of 
the quark and gluon structure of nucleons and nuclei. 
While initially nuclear effects in DIS were thought to be largely negligible, 
this was proven wrong by the measurement of the ratio of the
iron to deuterium structure functions performed by 
the European Muon Collaboration (EMC) at CERN in 1983~\cite{emc83}. 
The apparent
disagreement between the dramatic deviation of the ratio from unity seen in the 
EMC data and the small nuclear effects
predicted by theoretical calculations
has triggered
a series of further measurements and theoretical investigations, for reviews, 
see~\cite{emcreview_1,emcreview_2,emcreview_3,emcreview_4}.

\begin{figure}[h]
\centerline{\includegraphics[scale=0.45]{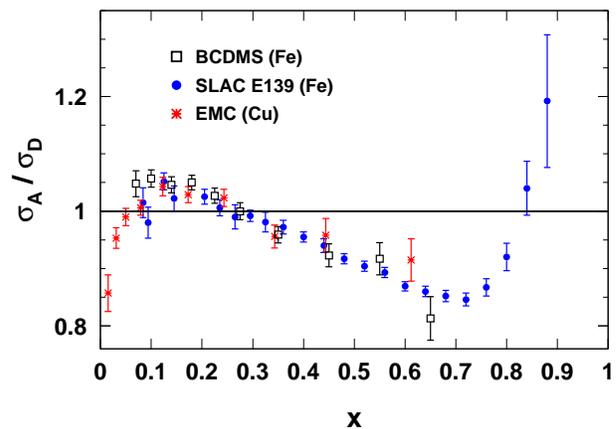}}
\caption{The pattern of nuclear modifications of the $\sigma^A/\sigma^D$ cross section ratio
as a function of Bjorken $x$ for $^{56}$Fe and $^{64}$Cu. The data are from
BCDMS~\cite{bcdms} (open squares), SLAC E139~\cite{e139} (filled circles) and
EMC~\cite{emc} (stars). For all data sets statistical and systematic errors have been combined 
in quadrature.}
\label{fig:emc_cu_fe}
\end{figure}

The emerging picture of nuclear modifications of the nucleus to deuteron 
cross section ratio, $\sigma^A/\sigma^D$,
has
the pattern presented in Fig.~\ref{fig:emc_cu_fe}. 
For small values of Bjorken $x$, 
$x < 0.05-0.1$, 
the ratio is noticeably suppressed---the suppression increases with an increase of the atomic 
number $A$ and a decrease of $x$---which is called nuclear shadowing. For $0.1 < x < 0.3$, 
the ratio is enhanced; the effect 
is small (of the order of a few percent) and does not reveal an obvious $A$ dependence. 
In the interval
$0.3<x<0.8$, the ratio is suppressed and this suppression is called the EMC effect.
Finally, for $x>0.8$ the ratio dramatically grows above unity which is explained by the effect 
of the nucleon motion inside nuclei (Fermi motion). 
Various models describe the experimental 
$\sigma^A/\sigma^D$ 
cross section ratios for certain ranges of Bjorken $x$, 
but there is no comprehensive understanding of the entire pattern of the 
nuclear modifications described 
above. In particular, there is no
unique and generally accepted theory to explain the nature of the 
antishadowing and EMC effects.

In this paper we focus on the enhancement (antishadowing) of the 
$\sigma^A/\sigma^D$ cross section ratios in the $0.1<x<0.3$ region. The
deviation of $\sigma^A/\sigma^D$ from unity in the antishadowing region is
of the order of a few percent~\cite{emcreview_1,emcreview_2,emcreview_3,emcreview_4} 
(see Fig.~\ref{fig:emc_cu_fe}). Given that most 
measurements 
quote normalization 
uncertainties on the order of 1-2\% (usually due to target thickness or 
luminosity), it is difficult to quantify the absolute size of the 
antishadowing effect 
precisely,
and comparisons 
between experiments are somewhat complicated. In addition, 
systematic uncertainties due to radiative corrections are highly non-trivial in
this region of $x$, and are sometimes hard to determine accurately. 
An example of the difficulty involved in achieving very precise measurements 
in the antishadowing region can be found in the SLAC E139 results. The 
preliminary results for the Fe/D ratio were essentially consistent with unity 
for the region $0.1<x<0.3$~\cite{e139_old}.  However, the final E139 analysis yielded 
results in the antishadowing region more consistent with, e.g., the EMC and 
BCDMS experiments, showing a small enhancement of 
$\approx 3\%$ on average~\cite{e139}. 
Despite the difficulties inherent in 
antishadowing measurements,
the 
results from various experiments are remarkably consistent within their 
experimental uncertainties. In addition, the small enhancement seen by the 
EMC, BCDMS, and SLAC E139 experiments for copper and iron targets has also been
seen in lighter targets (Ca/D, N/D, C/D, He/D) by the NMC~\cite{nmcca2d} and 
HERMES~\cite{Ackerstaff:1999ac} experiments. 

The antishadowing effect 
has rather intriguing features.  
Unlike the shadowing  effect, antishadowing showed little or no sensitivity to the mass number $A$
within experimental uncertainties, 
for example, in the  SLAC E139~\cite{e139} and
 NMC data~\cite{nmcca2d}. 
While antishadowing is observed in nuclear DIS,
the cross section enhancement is not seen in 
nuclear Drell-Yan rates~\cite{e772} and 
total neutrino-nucleus cross sections for $x>0.1$~\cite{iron08}.

In the leading twist formalism, the small enhancement of 
$\sigma^A/\sigma^D$
in the antishadowing region translates into an enhancement 
of the valence quark and 
possibly 
gluon distributions in nuclei in this 
region~\cite{iron08,Global_Fits_1,Global_Fits_2,Global_Fits_3,Global_Fits_4}. 
However, the pattern and especially the magnitude of nuclear modifications of 
the gluon distribution in nuclei
are very poorly constrained 
by present data.

The aim of this paper is to examine the impact of the  
nuclear dependence of $R=\sigma_L/\sigma_T$, i.e., 
the ratio of the longitudinal to transverse photoabsorption cross sections, 
on the extraction of the nucleus to deuteron structure function ratios, $F_2^A/F_2^D$ and
 $F_1^A/F_1^D$, from $\sigma^A/\sigma^D$ data. 
In particular, we demonstrate that 
in the presence of a small but non-zero difference between $R$ for nuclei and 
the nucleon, 
the nuclear enhancement
in the ratio of the transverse structure functions $F_1^A/F_1^D$
becomes significantly reduced (or even disappears in some cases),
indicating that antishadowing is dominated by the longitudinal contribution.
In addition, we analyze how the nuclear dependence of $R$ affects 
 the nuclear gluon distribtion
and emphasize the importance of measurements of $R$ in the DIS kinematics 
as a direct probe of the 
gluon distribution in nuclei.

\section{Nuclear dependence of $R$ and the ratio of nucleus and deuteron structure functions}

\subsection{Longitudinal contribution to the inclusive cross section}
\label{subsec:longitudinal}

In the one-photon exchange approximation, the spin-independent cross
section for inclusive electron scattering can be expressed as 
\begin{eqnarray}
\frac{d^2 \sigma}{d\Omega dE^{\prime}} &= &\Gamma\left[\sigma_T(x,Q^2) +
\epsilon \sigma_L(x,Q^2)\right]  \nonumber \\
&=& \Gamma \sigma_T(x,Q^2) \left[1 +\epsilon {R(x,Q^2)}\right] \,,
\label{eq:cs1}
\end{eqnarray}
where $\sigma_T$ ($\sigma_L$) is the cross section for
photoabsorption of purely transversely (longitudinally) polarized photons; $R=\sigma_L/\sigma_T$;
$\Gamma$ is the transverse virtual photon flux;
$\epsilon$ is the virtual photon polarization parameter.
 In the laboratory frame, the negative four-momentum squared 
(virtuality) of the exchanged photon is $-q^2=Q^2 =  4EE^{\prime}{\sin}^2 (\theta/2)$
and the Bjorken $x$ is $x=Q^2/[2M (E-E^{\prime})]$, where $E$ ($E^{\prime}$) is the energy of the
incident (scattered) electron, $\theta$ is the scattering angle, and $M$ is the nucleon mass.
The flux of transverse virtual photons can be expressed as 
$\Gamma = \alpha E^{\prime}(W^2 - M^2)/[4 \pi^2 Q^2 M E (1 - \epsilon)]$, where $\alpha$ is the 
fine structure constant and $W$ is the invariant energy of the virtual photon-proton system.
Finally, the virtual photon polarization parameter is:
\begin{eqnarray}
\epsilon & = & \left[1 + 2(1+\frac{\nu^2}{Q^2}) {\tan}^2 \frac{\theta}{2}\right]^{-1} \nonumber\\
& =& \frac{1-y-\frac{M^2 x^2 y^2}{Q^2}}{1-y+\frac{y^2}{2}+\frac{M^2 x^2 y^2}{Q^2}} \,.
\label{eq:epsilon}
\end{eqnarray}
where 
$\nu=E-E^{\prime}$; $y=\nu/E$. Note that in the second line of Eq.~(\ref{eq:epsilon}) we expressed $\epsilon$ in a Lorentz invariant form.

In terms of the 
structure functions $F_1(x,Q^2)$ and $F_2(x,Q^2)$ in the DIS region, 
the double
differential cross section can be written as
\begin{eqnarray}
\frac{d^2 \sigma}{d\Omega dE^{\prime}} & = & \Gamma \frac{4{\pi}^2 
\alpha}{x(W^2-M^2)} \nonumber \\
& \times & \left[ 2xF_1  + \epsilon \left( (1+\frac{4M^2 x^2}{Q^2}) F_2-2xF_1 \right) \right].
\label{eq:cs2}
\end{eqnarray}
A comparison of Eqs.~(\ref{eq:cs1}) and (\ref{eq:cs2}) shows that 
$F_1(x,Q^2)$
is purely transverse, while 
\begin{equation}
F_L(x,Q^2)= (1+\frac{4M^2 x^2}{Q^2}) F_2(x,Q^2)-2xF_1(x,Q^2)
\label{eq:fl}
\end{equation}
is purely longitudinal.  
Note that 
$F_2(x,Q^2)$ is a mixture of both the longitudinal and transverse contributions. Thus,
the ratio $R$ is 
\begin{equation}
R \equiv \frac{\sigma_L}{\sigma_T}=\frac{F_L(x,Q^2)}{2xF_1(x,Q^2)} \,.
\end{equation}

The nucleon 
structure function $F_2(x,Q^2)$ is proportional to the 
$d^2 \sigma/(d\Omega dE^{\prime})$
differential cross section in the  $\epsilon \rightarrow 1$ limit; 
it has been measured with high precision in various $x$ and $Q^2$ bins. 
The longitudinal structure function $F_L(x,Q^2)$, in contrast, is not measured 
as well as $F_2(x,Q^2)$: 
the data is sparse and imprecise for the proton and is 
even more limited
for nuclei. 
It is an experimental challenge to separate $F_2(x,Q^2)$ and $F_L(x,Q^2)$ which is 
usually done using the method of Rosenbluth separation, i.e., by measuring  
the cross section at different energies (at fixed $x$ and $Q^2$) to allow 
for a variation of $\epsilon$.

In this paper we shall use the parameterization of $R$ for the nucleon, $R^N$, that 
was obtained as the result of the global analysis of the SLAC hydrogen and deuterium
data~\cite{whitlow}. The same analysis also showed that $R^D=R^H$ to high accuracy, where
$R^D$ ($R^H$) refers to deuterium (hydrogen). 
An example of the values of $R^N$ in the kinematics used in this paper is presented 
in the middle panel of Fig.~\ref{fig:kinematics}.
Note that the more recent analysis of the SLAC E143 collaboration~\cite{Abe:1998ym} 
reported 
results for $R^N$
consistent with those of Ref.~\cite{whitlow}.

\subsection{Hints of nontrivial nuclear dependence of $R$}
\label{subsec:hints}

 Experimentally measured cross section ratios contain both transverse 
and longitudinal contributions of the structure functions. 
In terms of the structure function $F_2(x,Q^2)$, one can write the ratio of the nucleus 
to deuteron photoabsorption cross sections as
\begin{eqnarray}
\frac{\sigma^A}{\sigma^D} &=& \frac{F_2^A(x,Q^2)}{F_2^D(x,Q^2)} 
\frac{1+ R^D}{1+R^A}  \frac{1+\epsilon R^A}{1+\epsilon R^D} \nonumber\\
&\approx & \frac{F_2^A(x,Q^2)}{F_2^D(x,Q^2)} \left[1-\frac{\Delta R(1-\epsilon)}{(1+R^D) (1+\epsilon R^D)} \right] \,,
\label{eq:cs_F2}
\end{eqnarray}
where the superscript $A$ refers to the nucleus and 
the superscript $D$ refers to the deuteron; $\Delta R \equiv R^A-R^D$.
In the second line of Eq.~(\ref{eq:cs_F2}) 
we used the expansion in terms of the small
parameter  
$\Delta R$
and kept first two terms of the expansion.

Alternatively one can express the cross sections $\sigma^{A}$ and $\sigma^{D}$ in terms 
of the structure function  $F_1(x,Q^2)$ and obtain:
\begin{eqnarray}
\frac{\sigma^A}{\sigma^D} & = & \frac{F_1^A(x,Q^2)}{F_1^D(x,Q^2)}  \frac{1+\epsilon R^A}{1+\epsilon R^D} \nonumber\\ 
& = &  \frac{F_1^A(x,Q^2)}{F_1^D(x,Q^2)} \left[1+\frac{\epsilon \Delta R}{1+\epsilon R^D}  \right] \,.
\label{eq:cs_F1}
\end{eqnarray}
The cross section ratio $\sigma^A/\sigma^D$ can be identified with the structure function ratio
$F_2^A/F_2^D$ or $F_1^A/F_1^D$ only with the assumption of the trivial nuclear dependence 
of $R=\sigma_L/\sigma_T$, i.e., $R^A=R^D$, or in 
certain kinematic limits. 
In particular, $\sigma^A/\sigma^D=F_2^A/F_2^D$ at $\epsilon=1$ and 
$\sigma^A/\sigma^D=F_1^A/F_1^D$ at $\epsilon=0$.

\begin{figure}
\centerline{\includegraphics[width=3.in]{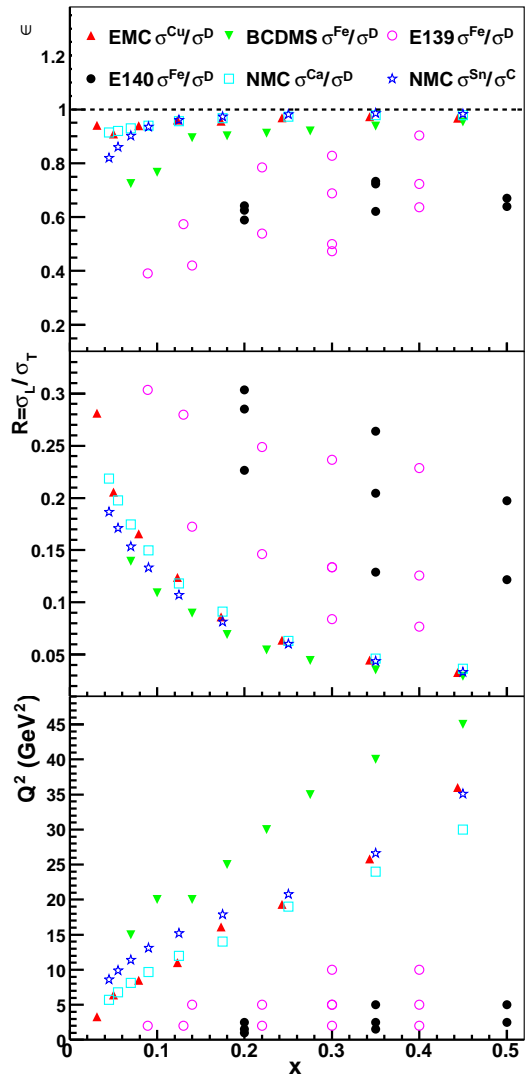}}
\caption{
 The kinematic coverage in $Q^2$ and $x$ and the 
corresponding values of $\epsilon$ and $R^N$ of the data points 
considered in this paper.
}
\label{fig:kinematics}
\end{figure}

Figure~\ref{fig:kinematics} presents the kinematic coverage in $Q^2$ and $x$ and the 
corresponding values of $\epsilon$ and $R^N$ of the data 
points 
considered in this paper.
On the one hand, 
the BCDMS~\cite{bcdms}, EMC~\cite{emc} and NMC~\cite{nmcca2d} data 
are mostly taken with $\epsilon$ close to unity (see the upper panel of Fig.~\ref{fig:kinematics}), 
which implies that the cross section ratios are close to the $F_2$ 
structure function ratios, 
even if $\Delta R \equiv R^A-R^D \neq 0$.
On the other hand, the SLAC data~\cite{e139,e140} 
(see the upper panel of Fig.~\ref{fig:kinematics}) 
corresponds to the kinematics  where 
 $\epsilon \approx 0.5$ and, hence, 
$F_2^A/F_2^D$ will deviate from  $\sigma^A/\sigma^D$ if $\Delta R \neq 0$.
For all these experiments $\epsilon \neq 0$ and, hence, the   
extraction of the  transverse structure function ratios 
$F_1^A/F_1^D$ 
depends 
explicitly 
on the assumption adopted
for 
 $\Delta R$.
Below we summarize what is known 
about it
from the world data.

At small $Q^2$, $R$ might be different between deuterium and hydrogen~\cite{e99118}, 
though it seems to be identical at large $Q^2$~\cite{nmcd2h,e140x}. 
In particular, 
there are some hints in both JLab E99-118~\cite{e99118} and SLAC data~\cite{e140x} 
that $R^D$ is smaller than  $R^H$ for $Q^2<1.5$ GeV$^2$, 
with a global average of $R^D-R^H=-0.054 \pm 0.029$.

Turning to heavier nuclei,
the SLAC E140 data~\cite{e140} suggest some nuclear dependence of $R$ at $x=0.2$, 
which seems to  have a nontrivial $Q^2$ dependence: $R^{Fe}-R^{D}$ can be positive at 
$Q^2=2.5$ GeV$^2$ and negative at $Q^2=1.5$ and $1$ GeV$^2$: \\
$R^{Fe}-R^{D} |_{Q^2=2.5} = 0.144 \pm 0.079 {\rm (stat.)}\pm0.027{\rm (syst.)}$; \\
$R^{Fe}-R^{D} |_{Q^2=1.5} = -0.124 \pm 0.051{\rm (stat.)} \pm0.023{\rm (syst.)}$:\\
 $R^{Fe}-R^{D} |_{Q^2=1} = -0.086 \pm 0.057{\rm (stat.)} \pm0.022{\rm (syst.)}$.
A word of caution is in order here.
Coulomb corrections may be non-negligible in DIS at SLAC 
and JLab kinematics, especially at large $x$.
These corrections tend to reduce $R$ for nuclear targets~\cite{solvignon_positron_wkshp}.

The nuclear dependence of $R$ at $Q^2$ of the order of a few GeV$^2$ and lower 
was also measured by the HERMES collaboration~\cite{Ackerstaff:1999ac} by 
fitting the $\sigma^A/\sigma^D$ cross section as a function of the virtual 
photon polarization $\epsilon$ over a typical range of $0.4 < \epsilon < 0.7$. 
Overall no significant nuclear dependence of $R$ for $^{14}$N and $^3$He targets for $x > 0.06$ 
has been observed (the data in the $x < 0.06$ region is affected by the correlated
background and should be neglected).
However, since this was a single-energy measurement with 
correlated  values of $\epsilon$ and
$Q^2$, the extraction of $R^A/R^D$ 
was done in a model-dependent way.

At larger values of $Q^2$, the NMC experiment~\cite{Amaudruz:1992wn} obtained
$R^{Ca}-R^{C}=0.027 \pm 0.026 ({\rm stat.}) \pm 0.020 (\rm syst.)$ at $\langle Q^2 \rangle =4$ GeV$^2$
and concluded that $\Delta R$ is compatible with zero.
However, a hint of the nontrivial nuclear dependence of $R$ can be still seen in data. 
The precision Sn/C data from NMC~\cite{nmcsn2c} show that $R^{Sn}-R^{C}=0.040 \pm 0.021 {\rm (stat.)} \pm 0.026 {\rm (syst.)}$ at a mean $Q^2$ of 10 GeV$^2$. 
This value of $\Delta R \equiv R^A-R^D$ corresponds to $\Delta R/R^N=0.22 \div 1.20$, 
i.e., $22 - 120$ \% relative deviation
for different values of 
$x$ in the considered 
kinematics, 
where $R^N$ is given by the parameterization of Ref.~\cite{whitlow} 
presented in the middle panel of Fig.~\ref{fig:kinematics}.
Note that the extraction of $\Delta R$ in this experiment was based on 
a method closely related to 
Rosenbluth separation taking advantage of three different incident 
muon energies (120, 200 and 280 GeV).

For convenience, we present in Table~\ref{table:RA_world_data} 
a brief overview (covered kinematics in Bjorken $x$ and
$Q^2$ and energy settings) of the 
discussed
measurements of the nuclear dependence of $R$ (involving 
nuclei heavier than deuterium). 
\begin{table}[h]
\begin{center}
\begin{tabular}{|c|c|c|}
\hline
Experiment and & Kinematics      &  Beam energy \\
observables    &                 &  \\
\hline
SLAC E140~\cite{e140} & $0.2 \leq x \leq 0.5$ & $3.7 \leq E \leq 15$ GeV, \\
$R^{Fe,Au}-R^D$                       & $1 \leq Q^2 \leq 10$ GeV$^2$ & up to five energies\\
\hline
NMC (1992)~\cite{Amaudruz:1992wn} & $0.0085 \leq x \leq 0.15$ & $E=90$ and 200 GeV,\\
$R^{Ca}-R^C$                      & $1 \leq Q^2 \leq 15$ GeV$^2$ & two energies   \\     
\hline           
HERMES~\cite{Ackerstaff:1999ac}   & $0.01 < x < 0.8$ & $E=27.5$ GeV \\
$R^{^3He,Ne}/R^D$                   & $0.2 < Q^2 < 3$ GeV$^2$  & single energy \\                    
\hline
NMC (1996)~\cite{nmcsn2c}         & $0.0125 \leq x \leq 0.45$ & $E=120$, 200, 280\\
$R^{Sn}-R^C$                       & $3.3 \leq Q^2 \leq 35$ GeV$^2$ & GeV, three energies \\
\hline
\end{tabular}
\caption{\label{table:RA_world_data}
An overview of the  measurements of the nuclear dependence of $R$ 
discussed in this paper.}
\end{center}
\end{table}

\begin{figure}
\centerline{\includegraphics[width=3.5in]{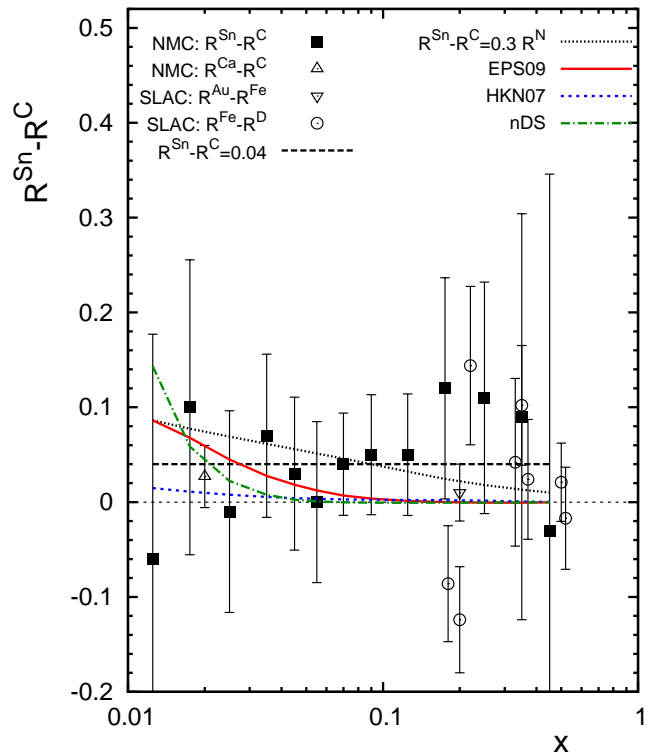}}
\caption{$R^{Sn}-R^{C}$ as a function of $x$. Full squares are results of the NMC 
measurement with the statistical and systematic errors added in quadrature~\cite{nmcsn2c};
the long-dash and dotted curves correspond to $R^{Sn}-R^{C}=0.04$ and 
$R^{Sn}-R^{C}=0.3 R^N$, respectively; 
the curves labeled ``EPS09'', ``HNK07'' and ``nDS'' correspond to predictions using 
different nuclear parton distributions. 
Also shown are the NMC result for $R^{Ca}-R^{C}$~\cite{Amaudruz:1992wn} (triangle), 
the SLAC result for $R^{Au}-R^{Fe}$~\cite{Dasu:1988ru} (inverse triangle), and 
SLAC E140 results for $R^{Fe}-R^{D}$ as a function of $x$~\cite{e140} (open circles).
}
\label{fig:Delta_R}
\end{figure}

The results of the NMC measurement of $R^{Sn}-R^{C}$ as a function of Bjorken $x$~\cite{nmcsn2c} 
are presented as
full squares in Fig.~\ref{fig:Delta_R}. For completeness, we also show 
the NMC result for the average $R^{Ca}-R^{C}$~\cite{Amaudruz:1992wn} as a triangle, 
the SLAC E140 result for the average $R^{Au}-R^{Fe}$~\cite{Dasu:1988ru} as an inverse  triangle, and
the SLAC E140 results for $R^{Fe}-R^{D}$ as a function of $x$~\cite{e140} as open circles.
(For the latter, we showed only the data points for the 6\% radiation length 
iron target and shifted
them in $x$ for better visibility.)
The long-dash and dotted curves correspond to $R^{Sn}-R^{C}=0.04$ and 
$R^{Sn}-R^{C}=0.3 R^N$, respectively. 
As one can see from the figure, 
both curves
provide a good description of the measured values of $R^{Sn}-R^{C}$.
Finally, the curves labeled ``EPS09'', ``HNK07'' and ``nDS'' correspond to the direct 
calculation of $R^{Sn}-R^{C}$ using different parameterizations of leading twist nuclear 
parton distributions (PDFs), see the discussion in Sect.~\ref{sec:discussion}.
Note that we have singled out the NMC Sn/C data~\cite{nmcsn2c} because 
the extraction of $R^A-R^D$ was done using a method closely related to the 
Rosenbluth separation and because 
the covered kinematics (the values of $x$, $Q^2$ and $\epsilon$) broadly overlaps with that
of the BCDMS, EMC, and NMC data on $\sigma^A/\sigma^D$ that we analyze in this paper.

In summary, 
as a global average, while $R$ seems to show 
little nuclear dependence within relatively large experimental uncertainties, 
there exist hints of nontrivial nuclear dependence of $R$. In particular, $\Delta R=R^A-R^D$ 
may be 
statistically different from zero
in some kinematics.

\subsection{Impact of nuclear dependence of $R$ on 
nucleus to deuteron structure function ratios}
\label{subsec:impact}

As we explained in Sect.~\ref{subsec:hints},
if there is a nontrivial nuclear dependence of $R$,
the $\sigma^A/\sigma^D$ cross section ratio is not equal to the
 $F_1^A/F_1^D$ or $F_2^A/F_2^D$  structure ratios.  
In particular, 
a positive $R^A-R^D$ will lead 
to $F_1^A/F_1^D < \sigma^A/\sigma^D <F_2^A/F_2^D $. 
Since the nuclear dependence of $R$ has not 
as yet 
been systematically measured, 
we 
shall
test two assumptions for $\Delta R$ that are 
motivated 
purely by the NMC Sn/C data~\cite{nmcsn2c}, which 
has  
kinematic coverage similar to that of the BCDMS, EMC and NMC 
measurements. 
In our analysis
below we assume that:\\
1) (Absolute) $\Delta R=R^A-R^D=0.04$. This is 
based on the NMC
measurement of $R^{Sn}-R^C$ at an average $\langle Q^2 \rangle=10$ GeV$^2$.
\\
2) (Relative) $(R^A-R^D)/R^N=30$\%, 
which is possible in view of the fact that the  NMC Sn/C data allows for
the $22 - 120$\% relative deviation of $\Delta R/R^N$.
 
Note that we effectively assumed that $R^A-R^D \approx R^{Sn}-R^{C}$ which corresponds
to the lower limit for $\Delta R$.

\begin{figure}
\centerline{\includegraphics[width=3.in]{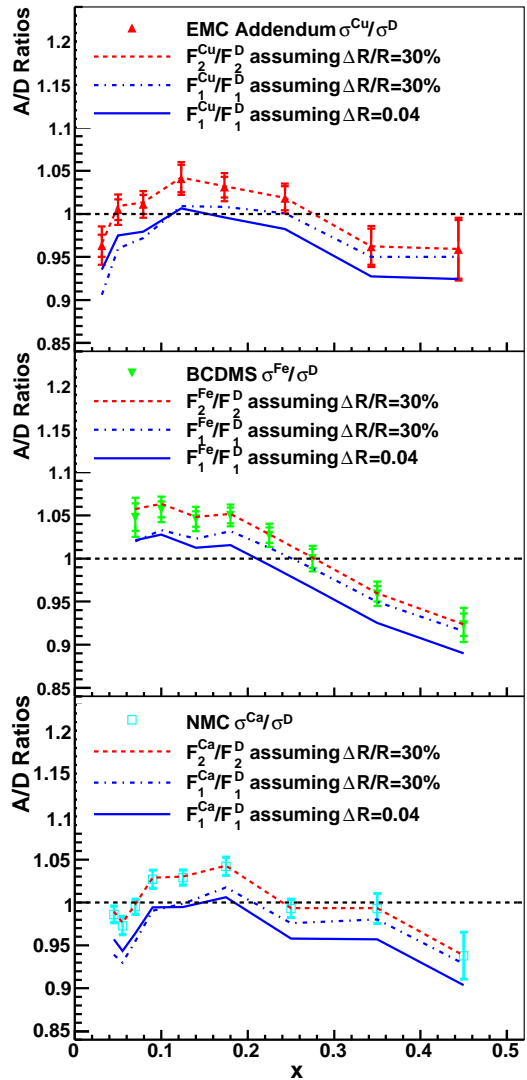}}
\caption{The impact of the nontrivial nuclear dependence of $R$ on the 
structure function ratios around the antishadowing region for BCDMS Fe/D~\cite{bcdms}, EMC Cu/D~\cite{emc} and NMC Ca/D~\cite{nmcca2d} data. The values of $\epsilon$ are close to unity.}
\label{fig:impact1}
\end{figure}

\begin{figure}
\centerline{\includegraphics[width=3.in]{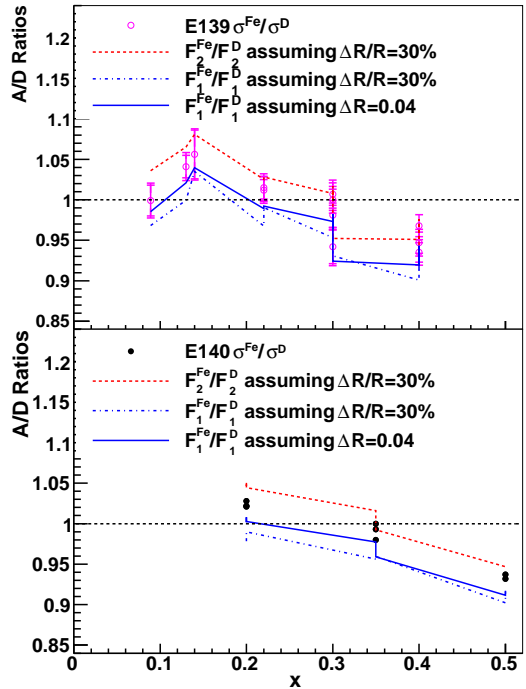}}
\caption{The impact of the nontrivial nuclear dependence of $R$ on 
the structure function
ratios around the antishadowing region for SLAC E139~\cite{e139} and E140~\cite{e140} Fe/D data.} 
\label{fig:impact2}
\end{figure}

The impact of our assumptions for 
$\Delta R$
on selected 
nuclear DIS data
 is presented in Figs.~\ref{fig:impact1} and \ref{fig:impact2},   
we truncated the used data sets by neglecting low $x$ and high $x$ data points
and focusing on the antishadowing region.

The BCDMS Fe/D~\cite{bcdms}, EMC Cu/D~\cite{emc} and NMC Ca/D~\cite{nmcca2d} data presented in
Fig.~\ref{fig:impact1} correspond to $\epsilon$ close to unity. Therefore, regardless of the 
assumption for $\Delta R$, one expects that $F_2^A/F_2^D \approx \sigma^A/\sigma^D$ with a very 
good accuracy. On the other hand, $F_1^A/F_1^D$ is clearly smaller than $\sigma^A/\sigma^D$.
Thus, the few percent enhancement of $\sigma^A/\sigma^D$  in the antishadowing region 
may be reduced or removed altogether for the ratio of the transverse structure
functions $F_1^A/F_1^D$.

 For the SLAC E139~\cite{e139} and E140~\cite{e140} Fe/D data presented in 
Fig.~\ref{fig:impact2}, the values of $Q^2$ are rather small (see the lower panel in
Fig.~\ref{fig:kinematics}) and our assumptions for the nuclear dependence of $R$ 
motivated by
 the NMC Sn/C measurement 
at higher $Q^2$ 
require a significant extrapolation in $Q^2$. 
However, for the lack of better input, in our analysis of the SLAC data
 we 
adopt
 the same assumptions for $\Delta R$ as those used above.
Since the values of $\epsilon$ for these 
two data sets are not close to unity (see the upper panel in Fig.~\ref{fig:kinematics}), 
$\Delta R > 0$ 
leads to noticeable differences between the ratio of the structure functions 
and the ratio of the cross sections according to the trend described by Eqs.~(\ref{eq:cs_F2})
and (\ref{eq:cs_F1}): $F_1^A/F_1^D < \sigma^A/\sigma^D <F_2^A/F_2^D$. 

In summary, 
the assumed
nontrivial nuclear dependence of $R$ leads to 
a decrease or to a complete disappearance (in some case) of enhancement of the 
 $F_1^A/F_1^D$ structure function ratio in the $0.1 < x < 0.3$ region.
If confirmed by future experiments, 
this observation would 
indicate that the effect of antishadowing in $\sigma^A$ 
is predominantly due to the contribution of the 
longitudinal structure function $F_L^A$, instead of $F_1^A$ as 
implicitly assumed in most phenomenological analyses.

\section{Experimental Limits on determining $R^A-R^D$}

Thus far we have examined the impact of a nuclear dependence of $R$ on the extraction 
of the nuclear dependent structure function ratios $F_1^A / F_1^D$ and $F_2^A / F_2^D$ 
from cross section ratios. 
The logical question then becomes: 'What is the limit on the experimental precision 
for $R^A-R^D$?'  
In this section we shall explore this question within the context of the precision likely to be 
avialable for dedicated L/T separation measurements over the next decade or two.  
For guidance we shall refer to the highest precision experiments performed 
at SLAC~\cite{e140,e140x,e139} 
and Jefferson Lab~\cite{e99118,Christy:2004rc}.
These experiments have shown that reducing the 
$\sigma_A / \sigma_D$
cross section ratio 
uncertainties, point-to-point in $\epsilon$, below 1\% is experimentally challenging, yet obtainable.  
For instance, the point-to-point uncertainties from Jefferson Lab experiment 
E94110~\cite{Christy:2004rc}
on cryogenic hydrogen have been estimated at about 1.5\%, which was found to be consistent 
with the width of the distribution of residuals determined from the linear fits.  

To measure cross sections at a range of $\epsilon$ values for fixed $x$ and $Q$, 
both the SLAC and JLab inclusive L/T separation experiments utilized a range of beam energies 
in conjunction with well studied spectrometer systems, which were able to be rotated to different 
angles and adjusted to accept varying ranges of momenta.  Some of the largest contributions 
to the estimated systematic uncertainties stem from either time dependent systematics, 
such as current calibrations or detector efficiency variations, or from the uncertainties 
in the kinematics at each beam energy and spectrometer setting.  
However, these systematics largely cancel in the cross section ratios, 
in which the electron yields on each target are taken at the same kinematic settings and close in time. 

If, for example, a 3\% anti-shadowing effect in $F_2^A$ were entirely due to a longitudinal enhancement, 
with $F_1^A / F_1^D = 1$, then this would be reflected in a 3\% slope in the cross section ratio 
versus $\epsilon^{\prime} \equiv \epsilon/(1+\epsilon R^D)$, 
corresponding to $\Delta R = R^A-R^D \approx 0.03$.  For the current study we assume the following:
\begin{itemize}
\item 
The total systematic point-to-point uncertainty (in $\epsilon$) on the measured 
$\sigma^A / \sigma^D$ ratios is 0.5\%.

\item 
There is no $\epsilon$ dependent systematic uncertainty.

\item 
Six cross section ratio measurements at equally spaced $\epsilon^{\prime} = \epsilon/(1+\epsilon R^D)$ 
values in the range (0.05, 0.95), corresponding to six unique beam energies.
\end{itemize}

\begin{figure}[t]
\begin{center}
\includegraphics[scale=0.45]{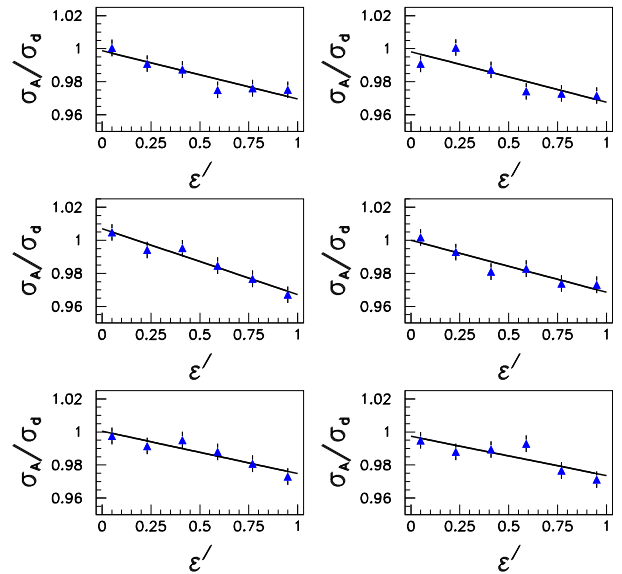}
\caption{Simulated $\sigma^A/ \sigma^D$ as a function of $\epsilon^{\prime}$ (data points with 
error bars) and a linear fit (solid lines), see the text for details. 
}
\label{fig:ra-rd-exp}
\end{center}
\end{figure}

Under the assumptions above, the cross section ratios were selected at each $\epsilon^{\prime}$ 
by random sampling from a Gaussian distribution assuming a 3\% slope on 
$\sigma^A/ \sigma^D$ versus $\epsilon^{\prime}$ and a Gaussian width of 0.5\%.  
Six sample L/T separations generated by this procedure are shown in Fig.~\ref{fig:ra-rd-exp}.  
After performing a linear fit, the uncertainty on the measured slope was found to be 0.67\%, 
corresponding to a 1-$\sigma$ (3-$\sigma$) uncertainty on $R^A-R^D$ of less than 0.007 (0.021).  
For the case considered of 0.5\% ratio uncertainties, one could determine at 1-$\sigma$ whether 
a 3\% antishadowing effect is due mainly to $F_L^A$ to  $\approx 20 \%$.  

We note that this uncertainty on the extracted $R^A-R^D$ scales with the uncertainties 
on the cross section ratios such that a further reduction in the latter to 0.25\% would 
reduce the uncertainty on $\Delta R$ by half.  However, we have thus far ignored any possible 
$\epsilon$ dependent systematic uncertainties, such as those possibly arising from Coulomb and 
radiative corrections.  For this reason, this is likely an optimistic scenario.

\section{Nuclear dependence of $R$ and its role in antishadowing 
of the gluon distribution in nuclei}
\label{sec:discussion}

We demonstrated in Sect.~\ref{subsec:impact} that the assumption 
of the nontrivial nuclear dependence
of $R$, i.e., $R^A-R^D > 0$, whose magnitude and sign are motivated by the NMC Sn/C 
data~\cite{nmcsn2c}, leads to a difference between the cross section and structure function 
ratios: $F_1^A/F_1^D < \sigma^A/\sigma^D <F_2^A/F_2^D$. 
Moreover, the reduction of the $F_1^A/F_1^D$ ratio is quite sizable: the enhancement 
in the $0.1 < x < 0.3$ region 
visible in the cross section ratios
is significantly decreased (or even disappears)
for the $F_1^A/F_1^D$ ratios,
which
indicates that antishadowing predominantly resides in the longitudinal structure function $F_L^A$.
This conclusion is rather general; 
in particular, it does not rely on the twist expansion and the underlying partonic structure.

\begin{figure}
\centerline{\includegraphics[scale=1.5]{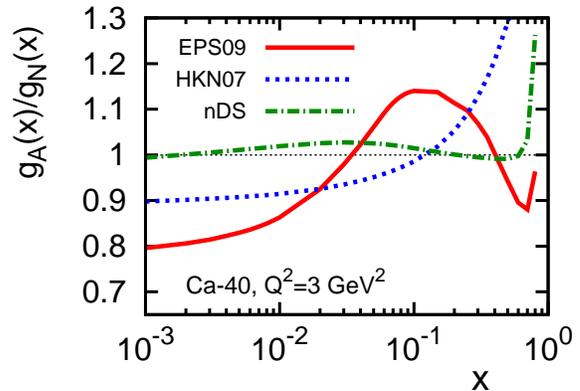}}
\caption{$g_A(x)/g_N(x)$ for $^{40}$Ca as a function of $x$ at fixed $Q^2=3$ GeV$^2$
as obtained from global QCD fits.
The solid (dotted, dot-dashed) curve is the result of the EPS09~\cite{Global_Fits_3}
(HKN07 fit~\cite{Global_Fits_2}, nDS~\cite{Global_Fits_1}) fit.}
\label{fig:ga_Ca40_Q2_3}
\end{figure}

In the framework of the leading twist formalism,  
global QCD fits to the available 
data~\cite{iron08,Global_Fits_1,Global_Fits_2,Global_Fits_3,Global_Fits_4} 
show that the small enhancement of 
$\sigma^A/\sigma^D$
in the antishadowing region translates into an enhancement of the valence quark and
possibly
gluon
distributions in nuclei 
compared to those in the free proton. 
One should emphasize that these analyses assumed no nuclear dependence of $R$, i.e.,
$R=0$.
The pattern and magnitude of nuclear modifications
of 
the nuclear gluon distribution
$g_A(x)$ is known with very large uncertainty because $g_A(x)$ is mostly 
determined 
 indirectly from scaling violations of the nuclear structure function
$F_{2}^A$ measured in a limited kinematics. This is illustrated in 
Fig.~\ref{fig:ga_Ca40_Q2_3} where we present the ratio of  
leading order 
gluon distributions in $^{40}$Ca to that in the free proton, $g_A(x)/g_N(x)$,
as a function of $x$ at fixed $Q^2=3$ GeV$^2$. In the figure,
the solid curve is the result of the EPS09 fit~\cite{Global_Fits_3}; the dotted curve 
is the result of the HKN07 fit~\cite{Global_Fits_2}; the dot-dashed curve is the 
nDS parameterization~\cite{Global_Fits_1},
whose results are quantitatively similar to those of~\cite{Global_Fits_4}.  
For the 
EPS09 and HKN07 fits, we showed only the central values; 
the theoretical uncertainty on these predictions is quite large essentially in the
entire range of $x$.

As one can see from Fig.~\ref{fig:ga_Ca40_Q2_3}, different groups predicts wildly different
$g_A(x)/g_N(x)$ (with large uncertainties). Since the amounts of nuclear shadowing and 
antishadowing are correlated through the momentum sum rule, 
large antishadowing corresponds to significant shadowing in the
EPS09 fit~\cite{Global_Fits_3}; very small antishadowing corresponds to 
negligibly small shadowing in the nDS fit~\cite{Global_Fits_1}); the HKN07 
fit~\cite{Global_Fits_2} suggests yet another scenario where large gluon antishadowing
is concentrated at large $x$.

Given the present uncertainty in $g_A(x)$, 
it is important and instructive to confront the NMC measurement of 
$R^{Sn}-R^{C}$~\cite{nmcsn2c} with direct calculations of 
this quantity in the framework of leading twist nuclear parton distributions.
This is presented in Fig.~\ref{fig:Delta_R} where the curves 
labeled ``EPS09'', ``HKN07'' and ``nDS'' correspond to the direct calculation 
of $R^{Sn}-R^{C}$ in the kinematics of the NMC measurement~\cite{nmcsn2c} using 
the respective
leading order parton distributions in nuclei.
One can readily see from the figure that while for small $x$, $x < 0.05$, 
the leading twist description is consistent with our assumptions for $\Delta R \neq 0$ 
and the NMC data,
in the antishadowing region $0.1 < x <  0.3$ and also for larger $x$ 
the leading twist approach predicts a negligibly small
$\Delta R$ in contrast with our assumptions and 
only marginally agrees with the data due to the large experimental 
uncertainty.
Note that the leading twist calculations presented in Fig.~\ref{fig:Delta_R}
have rather small theoretical uncertainties 
stemming mostly from the uncertainty in the gluon distributions.

\begin{figure}
\includegraphics[scale=1.5]{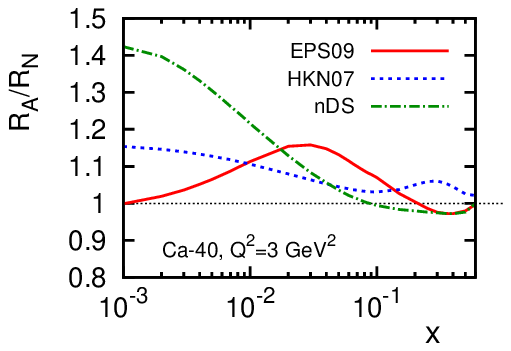}
\includegraphics[scale=1.5]{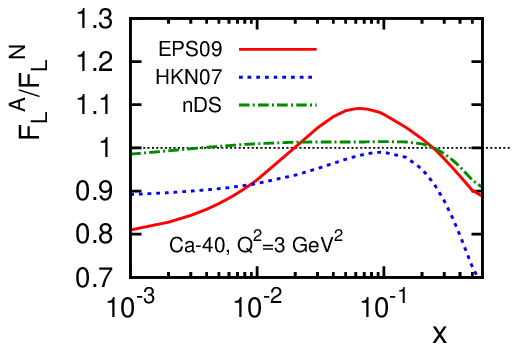}
\caption{$R^A/R^N$ (upper panel) and $F_L^A/F_L^N$ (lower panel)
$g_A(x)/g_N(x)$ for $^{40}$Ca as functions of $x$ at $Q^2=3$ GeV$^2$.
The solid (dotted, dot-dashed) curve is the result of the EPS09~\cite{Global_Fits_3}
(HKN07 fit~\cite{Global_Fits_2}, nDS~\cite{Global_Fits_1}) fit.}
\label{fig:RA_Ca40}
\end{figure}

There are several reasons for the negligibly small value of $R^{Sn}-R^{C}$ for $x > 0.05$ 
at the NMC energies predicted in the leading twist framework.
First and most importantly, the assumed shapes of the parameterizations of quark and gluon distribution in 
nuclei~\cite{Global_Fits_1,Global_Fits_2,Global_Fits_3,Global_Fits_4} are 
such that nuclear PDFs and the ratio $R$ show only a weak nuclear dependence 
around $x=0.1$ (see Figs.~\ref{fig:ga_Ca40_Q2_3} and \ref{fig:RA_Ca40}). 
For instance, 
while the EPS09 analysis~\cite{Global_Fits_3} used the  data on $Q^2$ dependence 
of $F_2^{Sn}(x,Q^2)/F_2^{C}(x,Q^2)$~\cite{nmcsn2c}, it did not include the 
$R^{Sn}-R^{C}$ data in the fit. Hence, resulting nuclear PDFs were not constrained 
to reproduce the experimental values of $R^{Sn}-R^{C}$ which, as a result, leads to 
$R^{Sn}-R^{C} \approx 0$ for $x > 0.1$.
Second, while $R^{Sn}/R^N$ and $R^{C}/R^N$ separately reveal quite sizable deviations 
from unity (compare to $R^{Ca}/R^N$ presented in the upper panel of Fig.~\ref{fig:RA_Ca40}),
nuclear effects mostly cancel in the $R^{Sn}-R^{C}$ difference. 
In general, while it is natural to expect $\Delta R \neq 0$ 
because the pattern of nuclear modifications of quark and gluon distributions 
is different,  
with the currently assumed shapes of nuclear parton distributions 
it is not easy to generate sizable $\Delta R$ for $x > 0.1$ and large $Q^2$ 
because $R$ itself is 
very small there.
Third, in the NMC kinematics
the values of Bjorken $x >0.1$ correspond to $Q^2 > 10$ GeV$^2$. At such large values of 
$Q^2$, nuclear modifications of parton distributions gradually become less pronounced.
Note also that it is unlikely that higher twist (twist-four) effects can 
generate sizable $\Delta R$
because it would require unrealistically large higher twist effects~\cite{emcreview_1}.

While the available data on the nuclear dependence of $R$ is not able to 
constrain the nuclear gluon distribution in the $0.1 < x < 0.3$ region, 
a better chance of measuring gluon antishadowing would be offered by 
measurements of $R$ with 
nuclear targets and the deuteron (proton) and 
at not too high $Q^2$. Note this is  essentially equivalent 
to measuring the longitudinal structure functions $F_L(x,Q^2)$ for nuclei and the deuteron 
(proton). Such measurements can be carried out at Jefferson Lab 
at 12 GeV at low-to-intermediate $Q^2$~\cite{pr1211113} 
and at a future Electron-Ion Collider (EIC) at intermediate-to-high 
$Q^2$~\cite{Boer:2011fh,Accardi:2011mz}.  
In the latter case, the measurement of $F_2^A(x,Q^2)$ and
the longitudinal nuclear structure function 
$F_L^A(x,Q^2)$ (taking advantage of variable energies)
with the subsequent extraction of $g_A(x)$ in a wide kinematic range is 
already 
an 
important part of the planned physics program.

An example of expected nuclear effects is presented in
Fig.~\ref{fig:RA_Ca40} which shows predictions for the ratio of the nuclear to nucleon
ratios $R^A/R^N$ (upper panel) and longitudinal structure functions $F_L^A/F_L^N$ 
(lower panel) as a function 
of $x$ at $Q^2=3$ GeV$^2$ for $^{40}$Ca. Different curves correspond to 
different parameterizations of nuclear PDFs (see Fig.~\ref{fig:ga_Ca40_Q2_3}).
A comparison of Figs.~\ref{fig:ga_Ca40_Q2_3} and \ref{fig:RA_Ca40} shows that different
assumptions about the shape of the gluon (and quark) distributions in nuclei lead to
different shapes of $R^A/R^N$ and $F_L^A/F_L^N$.
To point out just one feature, an observation of sizable $R^A/R^N > 1$ 
(enhanced $F_L^A/F_L^N$ compared to $F_1^A/F_1^N$) and $F_L^A/F_L^N > 1$
in the antishadowing region 
$0.1 < x < 0.3$
 would unambiguously signal the presence 
of a significant antishadowing for the gluon distribution in nuclei.
(The gluon distribution enters the longitudinal 
structure function $F_L(x,Q^2)$ at the same
order as the quark distributions; at the same time, the gluon distribution 
enters the transverse structure function 
$F_1(x,Q^2)$ with the weight (coefficient function) that is smaller than that for 
$F_2(x,Q^2)$~\cite{Brock:1993sz}.)
The converse is also true:
an absence of nuclear enhancement of $R^A/R^N$ and  $F_L^A/F_L^N$ in the interval 
$0.1 < x < 0.3$
would translate
into the absence of antishadowing for gluons in this region.

\section{Conclusions}

In this paper we studied 
the influence of the nontrivial nuclear dependence of $R=\sigma_L/\sigma_T$ 
on 
the extraction of the $F_2^A/F_2^D$ and $F_1^A/F_1^D$ structure function ratios 
from the data on the 
$\sigma^A/\sigma^D$ cross section ratios.  
Guided by indications of the nuclear dependence of $R$ from the world data and, in particular, by
the NMC measurement 
that showed that
$R^{Sn}-R^{C}=0.040 \pm 0.021 {\rm (stat.)} \pm 0.026 {\rm (syst.)}$ at $\langle Q^2 \rangle = 10$ 
GeV$^2$~\cite{nmcsn2c},
we tested two assumptions for $\Delta R \equiv R^A -R^D$: $\Delta R=0.04$ and $\Delta R/R^N=0.3$, where 
$R^N$ corresponds to the free proton~\cite{whitlow}.
With these assumptions, we examined selected sets of EMC, BCDMS, NMC and SLAC data on 
$\sigma^A/\sigma^D$ and extracted the $F_2^A/F_2^D$ and $F_1^A/F_1^D$ ratios. We find that
for the  EMC, BCDMS and NMC data,   $F_2^A/F_2^D \approx \sigma^A/\sigma^D$, while 
$F_1^A/F_1^D <\sigma^A/\sigma^D$. For the SLAC data, we found that 
$F_1^A/F_1^D <\sigma^A/\sigma^D < F_2^A/F_2^D$. In particular, we observed that 
the nuclear enhancement (antishadowing) 
in the interval $0.1 < x < 0.3$ 
becomes significantly reduced (or even disappears in some cases) 
for the ratio of the transverse structure functions $F_1^A/F_1^D$.
The latter observation indicates that antishadowing 
may in fact be dominated by the longitudinal contribution rather than by the 
transverse one (i.e., antishadowing is dominated by gluons rather than by quarks) 
as implicitly assumed by current phenomenological analyses and global nuclear  
parton distribution fits.

We also examined experimental limits on determining $R^A-R^D$ from measurements of
the $\epsilon^{\prime}=\epsilon/(1+\epsilon R^D)$ dependence of $\sigma^A/\sigma^D$.
Making a plausible assumption that $\sigma^A/\sigma^D$ has a 3\% slope in $\epsilon^{\prime}$ 
and  can be measured with a 0.5\%
uncertainty over a broad range of $\epsilon^{\prime}$, we found that $\Delta R$ can be 
extracted with 0.67\% uncertanity. 
Therefore, one could determine whether a 3\% antishadowing effect is mainly due to $F_L^A$ to
approximately 20\% accuracy.

In the leading twist framework, the magnitude of 
nuclear enhancement of $R^A$ and the longitudinal structure function $F_L^A(x,Q^2)$ 
(these quantities directly probe 
the nuclear gluon distribution $g_A(x)$) 
is directly correlated with the size and shape of antishadowing for $g_A(x)$.
While at the moment $g_A(x)$ is rather poorly constrained by QCD fits to available data, 
a dedicated high-precision measurement of the nuclear dependence of $R$ 
(the longitudinal nuclear structure function
$F_L^A(x,Q^2)$) at Jefferson Lab and an EIC  
has the potential to unambiguously constrain $g_A(x)$ in the antishadowing region and beyond
(An EIC will also be able to constrain $g_A(x)$ deep in the shadowing region of small $x$.) 
Through the parton momentum sum rule, this knowledge will have a deep impact on  $g_A(x)$ 
in the entire range of $x$. In particular, it should dramatically help to constrain $g_A(x)$ 
in the nuclear shadowing region, $10^{-5} \leq x < 0.05$, where $g_A(x)$ plays an essential role 
in phenomenology of high-energy hard processes with nuclei, for a review, see~\cite{Frankfurt:2011cs}.

\section*{Acknowledgments}

We are grateful to J. Gomez, W. Melnitchouk, P. Monaghan and M. Strikman for helpful discussions.
This work was supported by the US Department of Energy contract No. DE-AC05-06OR23177, under which Jefferson Science Associates, LLC operates Jefferson Lab, and US National Science Foundation award No. 1002644.

\end{document}